\documentclass[prb,twocolumn,showpacs,floatfix,amsmath,amssymb,superscriptaddress]{revtex4-1}
\usepackage{amsfonts}
\usepackage{stmaryrd}
\usepackage{bbm}
\usepackage{mathrsfs}
\usepackage{tipa}
\usepackage{amssymb}
\usepackage{txfonts}
\usepackage{graphicx}
\usepackage{dcolumn}
\usepackage{epstopdf}
\usepackage[colorlinks,linkcolor=blue,urlcolor=blue,citecolor=blue]{hyperref}
\usepackage{multirow}
\usepackage{subfigure}
\usepackage{url}

\begin{document}

\title{Identifying and tracking magnetically induced polarization in Fe$_2$Mo$_3$O$_8$ by static and time-resolved second harmonic generation}
\author{L. Y. Shi}
\affiliation{International Center for Quantum Materials, School of Physics, Peking University, Beijing 100871, China}

\author{D. Wu}
\affiliation{Beijing Academy of Quantum Information Sciences, Beijing 100913, China}
\affiliation{Songshan Lake Materials Laboratory, Dongguan, Guangdong 523808, China}

\author{T. Lin}
\affiliation{International Center for Quantum Materials, School of
Physics, Peking University, Beijing 100871, China}

\author{S. J. Zhang}
\affiliation{International Center for Quantum Materials, School of
Physics, Peking University, Beijing 100871, China}
\affiliation{Beijing National Laboratory for Condensed Matter Physics, Institute of Physics, Chinese Academy of Sciences, Beijing 100190, China}

\author{Q. M. Liu}
\affiliation{International Center for Quantum Materials, School of
Physics, Peking University, Beijing 100871, China}

\author{Z. X. Wang}
\affiliation{International Center for Quantum Materials, School of
Physics, Peking University, Beijing 100871, China}

\author{T. C. Hu}
\affiliation{International Center for Quantum Materials, School of
Physics, Peking University, Beijing 100871, China}

\author{T. Dong}
\affiliation{International Center for Quantum Materials, School of Physics, Peking University, Beijing 100871, China}
\affiliation{Tsung-Dao Lee Institute, Shanghai JiaoTong University, Shanghai 201210, China}

\author{N. L. Wang}
\email{nlwang@pku.edu.cn}
\affiliation{International Center for Quantum Materials, School of Physics, Peking University, Beijing 100871, China}
\affiliation{Beijing Academy of Quantum Information Sciences, Beijing 100913, China}
\affiliation{Tsung-Dao Lee Institute, Shanghai JiaoTong University, Shanghai 201210, China}

\begin{abstract}
 Multiferroic materials offer a promising platform for ultrafast optical control of coupled magnetic and polar orders. However, a prerequisite for such control is to precisely identify how the magnetically induced polarization manifests itself on the ultrafast timescale, and then to probe its dynamics upon external perturbations. Here, we address this issue in the polar magnet Fe$_2$Mo$_3$O$_8$ by combining static and time-resolved second harmonic generation (SHG). Temperature-dependent static SHG reveals that, among the symmetry-allowed tensor elements, only $\chi^{(2)}_{ccc}$ exhibits a pronounced anomaly at the antiferromagnetic transition ($T_{\rm N} \approx 60$ K), identifying the $c$-axis polar response as the primary degree of freedom coupled to the magnetic order. Guided by this result, time-resolved SHG selectively tracks the dynamics of this tensor element following ultrafast photoexcitation. We observe a rapid enhancement of the $\chi^{(2)}_{ccc}$-related SHG signal, followed by biexponential recovery. The response is independent of the pump polarization, consistent with an ultrafast thermal origin, but is achieved at a fluence significantly below that required for conventional lattice heating. These results establish SHG as tensor-selective probe of ultrafast magnetoelectric dynamics and demonstrate the high sensitivity of the magnetically induced polarization in Fe$_2$Mo$_3$O$_8$ to optical excitation.
\end{abstract}

\maketitle

Multiferroic materials with coupled ferroelectricity and magnetism have attracted much attention due to their potential for mutual control of magnetization and polarization via electric and magnetic fields \cite{Kimura,TOKURA20071145,Fiebig2016}, and also provide a platform for ultrafast optical manipulation of different orders. Among optical techniques, time-resolved Kerr rotation and time-resolved second harmonic generation (SHG) are two important probes. Kerr rotation is widely used to study ferromagnetic materials \cite{PhysRevLett.76.4250,PhysRevLett.85.1986,PhysRevLett.94.087202,critical1}, but its application to antiferromagnets (AFMs) is challenging due to the lack of net magnetization. In contrast, SHG is intrinsically sensitive to inversion symmetry breaking, making it a powerful tool for probing electric polarization and lattice structure. Moreover, the breaking of time-reversal symmetry in magnetically ordered systems can also contribute to the SHG response. As a result, SHG has been extensively employed to investigate noncentrosymmetric AFMs and multiferroics \cite{PhysRevLett.84.5620,app8040570,PhysRevB.79.140411,PhysRevB.98.100301,Cheong2019,Ron2019,PhysRevLett.124.027601}. However, because the measured SHG signal generally contains contributions from multiple symmetry-allowed nonlinear susceptibility tensor elements, identifying the component associated with the magnetically induced polarization is essential for interpreting its ultrafast dynamics.

\begin{figure}[htbp]
	\centering
	\includegraphics[width=7cm]{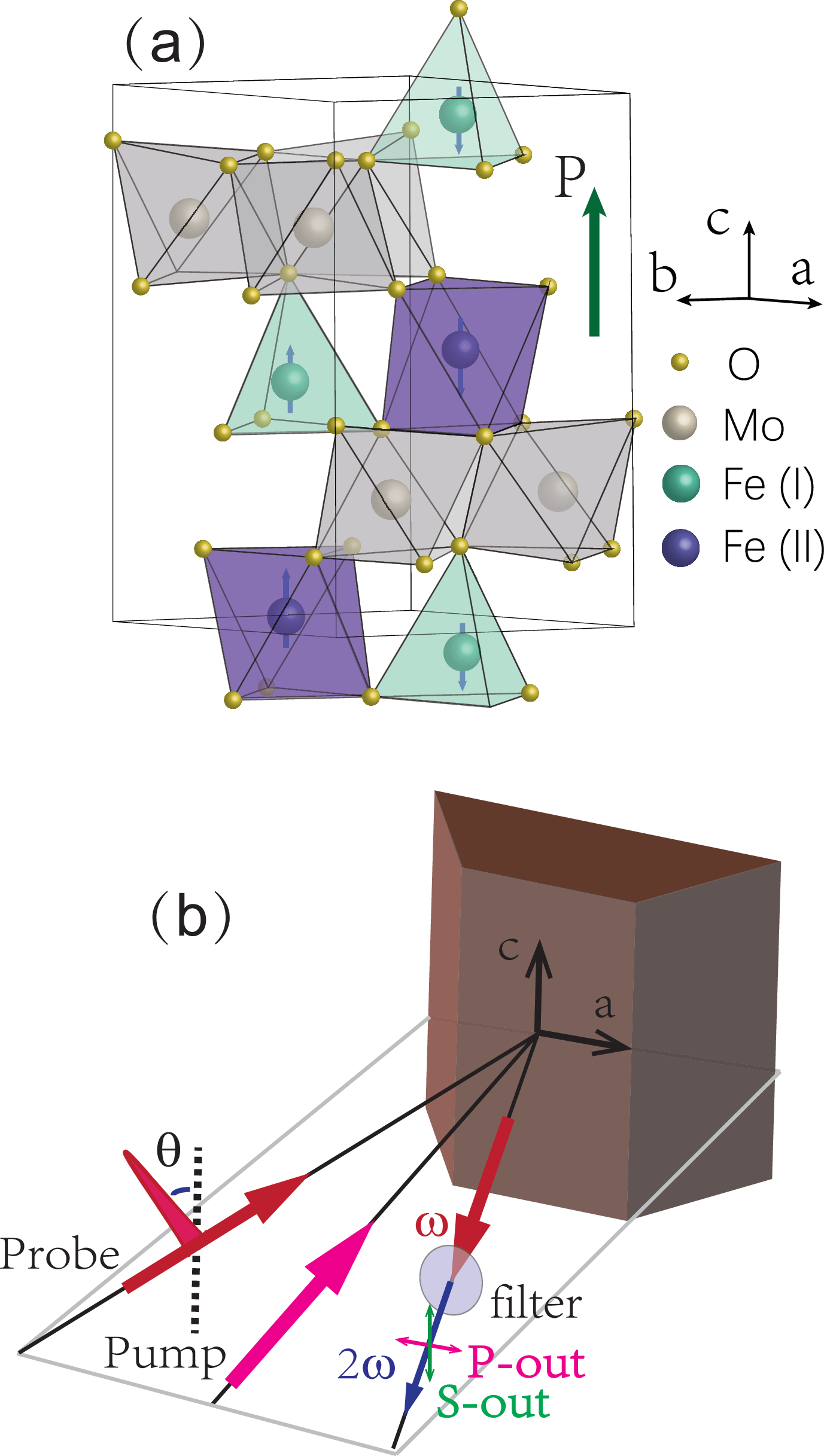}\\
	\caption{(a) Crystal and magnetic structure of Fe$_{2}$Mo$_{3}$O$_{8}$.  (b) Schematic illustration of measurement configuration of the TR-SHG.}\label{Fig:1}
\end{figure}

Fe$_2$Mo$_3$O$_8$ is a polar magnet that has attracted considerable interest \cite{LePage:a21124,Wang2015,PhysRevX.5.031034,0295-5075-118-3-37001,PhysRevB.95.020405,AxionResponse,0295-5075-118-3-37001,Roomteperaruremagnon,PhysRevX.9.031038,doi:10.1063/5.0044565,PhysRevB.102.174407,PhysRevB.102.094307,PhysRevB.102.115139, gcgl-9sbb, pq5m-32wj}. Its structure, illustrated in Fig. \ref{Fig:1}(a), consists of alternating honeycomb Fe layers and Kagome Mo layers stacked along the $c$-axis. 
The lattice belongs to the space group P6$_3$mc, with a spontaneous polarization along the $c$-axis \cite{doi:10.1021/ja01577a021}. Below $T_N = 60$ K, the Fe spins exhibit antiferromagnetic order along the $c$-axis \cite{doi:10.1021/ic50019a003,MCALISTER1983340}. The magnetic ordering is accompanied by a structural distortion: oxygen atoms shift along the $c$-axis, giving rise to an additional polarization change $\Delta P$ and an increase of the Fe-O-Fe bond angle, which enhances the in-plane exchange coupling $J$ \cite{Wang2015}. These lattice modifications are attributed to exchange striction \cite{Wang2015,PhysRevX.5.031034,PhysRevLett.131.136701,PhysRevB.109.094419}. Subsequent studies have further uncovered a rich variety of magnetoelectric phenomena including topological magnon polarons and giant phonon magnetic moments\cite{Bao_2023,Wu_2023}.  In addition, a strong magneto-optical Kerr effect has been observed under circularly polarized pumping \cite{PhysRevX.9.031038}. These characteristics make Fe$_2$Mo$_3$O$_8$ an ideal platform for identifying the nonlinear optical response associated with the magnetically induced polarization and tracking its ultrafast dynamics.

Here, we combine static and time-resolved SHG in a two-step approach. First, temperature-dependent static SHG measurements reveal a prominent reduction in only one of the second-order tensor elements, $\chi_{ccc}^{(2)}$, below the Néel temperature $T_N=60$ K, identifying $\chi_{ccc}^{(2)}$ as the nonlinear susceptibility tensor element coupled to the magnetically induced polarization. Guided by this identification, we then use time-resolved SHG to track the ultrafast dynamics of this tensor element following photoexcitation. The $\chi_{ccc}^{(2)}$-related SHG signal exhibits an ultrafast enhancement within $\sim300$ fs, reaching saturation at a remarkably low fluence of $\sim0.4$ mJ/cm$^2$, followed by a biexponential recovery. This work demonstrates that combining static and time-resolved SHG provides a tensor-selective strategy for investigating ultrafast magnetoelectric dynamics in multiferroic materials.

The Fe$_2$Mo$_3$O$_8$ single crystal was grown by the chemical vapor transport method \cite{STROBEL1983329,STROBEL1982242,PhysRevB.75.155406}. Static and time-resolved SHG measurements were performed on the natural ac-surface of the sample using a standard pump-probe system, with the experimental geometry illustrated in Fig. \ref{Fig:1}(b). The light source was a Ti:sapphire amplifier delivering 800 nm, 35 fs pulses at 1 kHz repetition. The pump beam was normally incident on the sample, while the probe beam was reflected from the sample at a small incident angle. The polarization of the incident probe was controlled by a half-wave plate, and the SHG signal was detected by a photomultiplier tube after filtering out the fundamental light. The pump spot size was $\sim 120$ $\mu$m and the probe spot size was $\sim 90$ $\mu$m. The fluence of the probe beam was $\sim 0.2$ mJ/cm$^2$. For temperature-dependent measurements, the sample was mounted in a closed-cycle cryostat.


In magnetic materials, the second-order nonlinear polarization can be expressed as $P_i(2\omega)= \epsilon_0\sum_{j,k}(\chi_{ijk}^{i(2)}+\chi_{ijk}^{c(2)})E_j(\omega)E_k(\omega)$, where $\chi_{ijk}^{i(2)}$ and $\chi_{ijk}^{c(2)}$ are the time-invariant and time-noninvariant nonlinear tensors, respectively \cite{Fiebig:05,PhysRevB.79.140411}. For Fe$_2$Mo$_3$O$_8$ in the paramagnetic state (space group P6$_3$mc, point group 6mm), the electric-dipole SHG tensor has only two independent nonzero components: $\chi_{ccc}^{(2)}$ and $\chi_{caa}^{(2)}=\chi_{aca}^{(2)}$, both of which are accessible in our ac-plane geometry. Below $T_N$, the magnetic point group becomes $\underline{6}m\underline{m}$, which allows nonzero $c$-type tensor components ($\chi_{bbb}^{c(2)}=-\chi_{baa}^{c(2)}=-\chi_{aba}^{c(2)}=-\chi_{aab}^{c(2)}$). However, for near-normal incidence on the ac-plane, these components are not detectable. We also checked an ab-cut sample and found no detectable signal at the magnetic transition, confirming that the magnetic-dipole contribution is negligible compared to the electric-dipole SHG.

\begin{figure}[htbp]
	\centering
	\includegraphics[width=8cm]{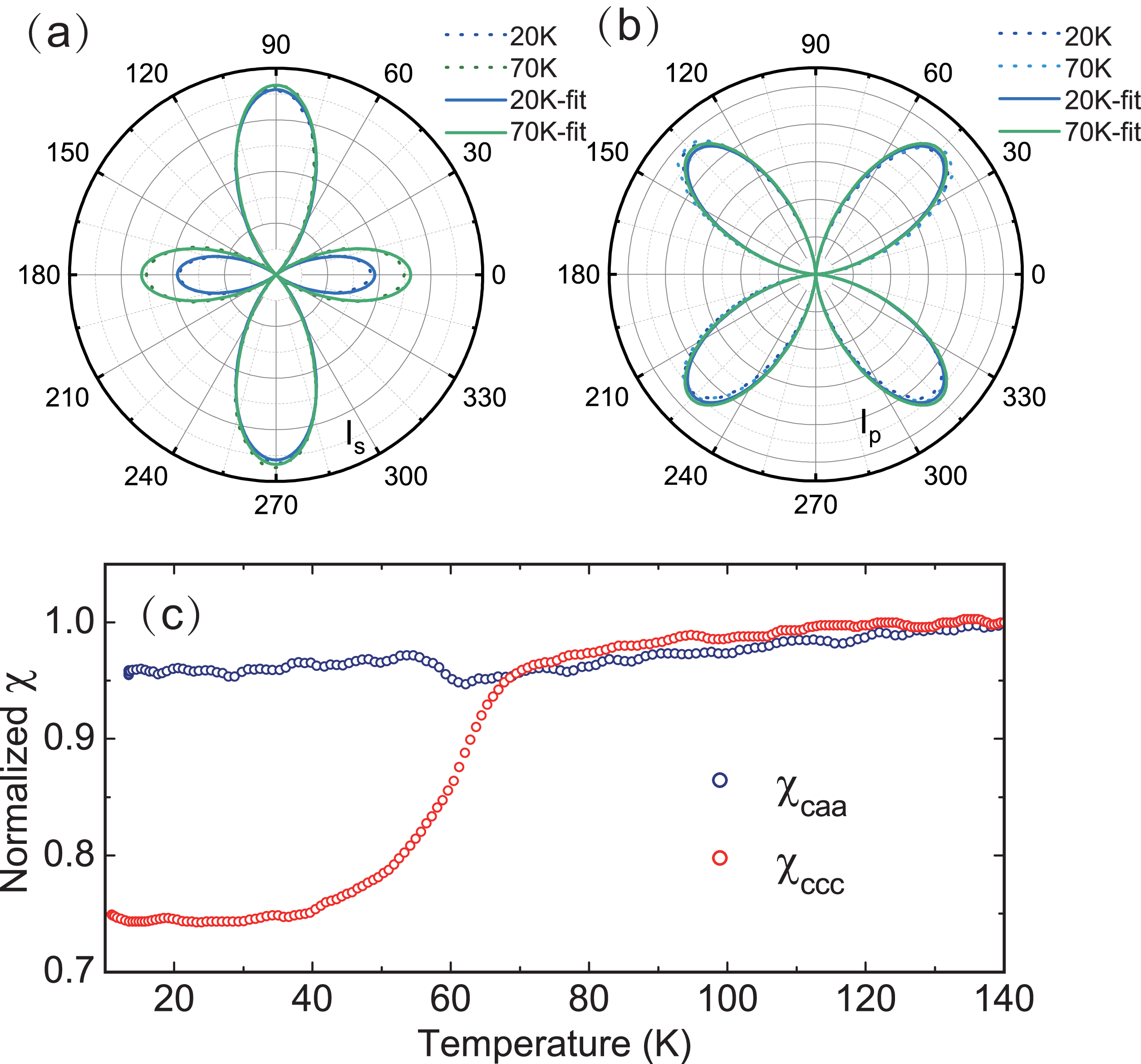}
	\caption{(a), (b) Polar plots of the SHG signal for s-out and p-out polarizations at 20 K and 70 K. The solid lines are fits to the expressions described in the text. (c) Temperature dependence of the SHG signals corresponding to $\chi_{ccc}^{(2)}$ and $\chi_{caa}^{(2)}$, normalized at 140 K. A clear anomaly is observed for $\chi_{ccc}^{(2)}$ near $T_N = 60$ K, while $\chi_{caa}^{(2)}$ shows no significant change.}\label{Fig:2}
\end{figure}

Figures \ref{Fig:2}(a) and (b) show the polar-angle-dependent SHG signals for s-out and p-out configurations at 20 K and 70 K. The SHG intensity can be expressed as $I_{s}\propto |\chi_{ccc}^{(2)}\cos^{2}\theta +\chi_{caa}^{(2)}\sin^{2}\theta|^{2}$ and $I_{p}\propto |2\chi_{aca}^{(2)}\sin2\theta|^{2}$ (see the Supplemental Material for derivation). The $\theta$-dependent patterns are well reproduced by these expressions, as shown by the solid lines in Fig. \ref{Fig:2}. From the fitting to $I_{s}$, we find that $\chi_{ccc}^{(2)}$ and $\chi_{caa}^{(2)}$ have opposite signs, and $|\chi_{caa}^{(2)}| > |\chi_{ccc}^{(2)}|$. As the temperature decreases from 70 K to 20 K, a prominent reduction of $I_s$ is observed only in the lobe near $\theta = 0^\circ$ (where $I_s$ is dominated by $\chi_{ccc}^{(2)}$), while the other lobe and all $I_p$ lobes remain unchanged. The polarization-resolved measurements therefore establish $\chi_{ccc}^{(2)}$ as the only nonlinear susceptibility component that is directly affected by the magnetic transition. This identification provides the basis for the time-resolved measurements presented below.

We now turn to the temperature evolution of individual tensor elements. The SHG signal related to $\chi_{ccc}^{(2)}$ is obtained with s-in ($E^\omega \parallel c$) and s-out ($E^{2\omega} \parallel c$) polarization configuration. For comparison, $\chi_{caa}^{(2)}$ and $\chi_{aca}^{(2)}$ are measured using p-in/s-out and p-out at $\theta = 45^\circ$, respectively. Figure \ref{Fig:2}(c) shows the temperature dependence of $\chi_{ccc}^{(2)}$. In the paramagnetic state, the SHG intensity decreases slightly upon cooling from 140 K, which is attributed to lattice anharmonicity. At $T_N$, a sharp drop of about 20\% is observed in $\chi_{ccc}^{(2)}$, while $\chi_{caa}^{(2)}$ shows no significant change across the transition (apart from a barely visible kink that is negligible compared to the drop in $\chi_{ccc}^{(2)}$).

The electric-dipole SHG signal is generally considered to be proportional to the electric polarization. In Fe$_2$Mo$_3$O$_8$, a magnetically induced polarization modulation $\Delta P$ along the $c$-axis has been established at $T_N$ \cite{Wang2015,PhysRevX.5.031034}. The reduction of $\chi_{ccc}^{(2)}$ therefore indicates that the magnetically induced polarization opposes the primal lattice polarization, meaning that the magnetic ordered state has a smaller lattice distortion along the $c$-axis than the paramagnetic state. Notably, only $\chi_{ccc}^{(2)}$ changes with the polarization, while all other components remain unaffected. This selective response may be understood from the fact that the exchange-striction-induced $\Delta P$ is directed along the $c$-axis \cite{Wang2015}, and thus couples predominantly to the $\chi_{ccc}^{(2)}$ channel.


\begin{figure}[htbp]
	\centering
	\includegraphics[width=7cm]{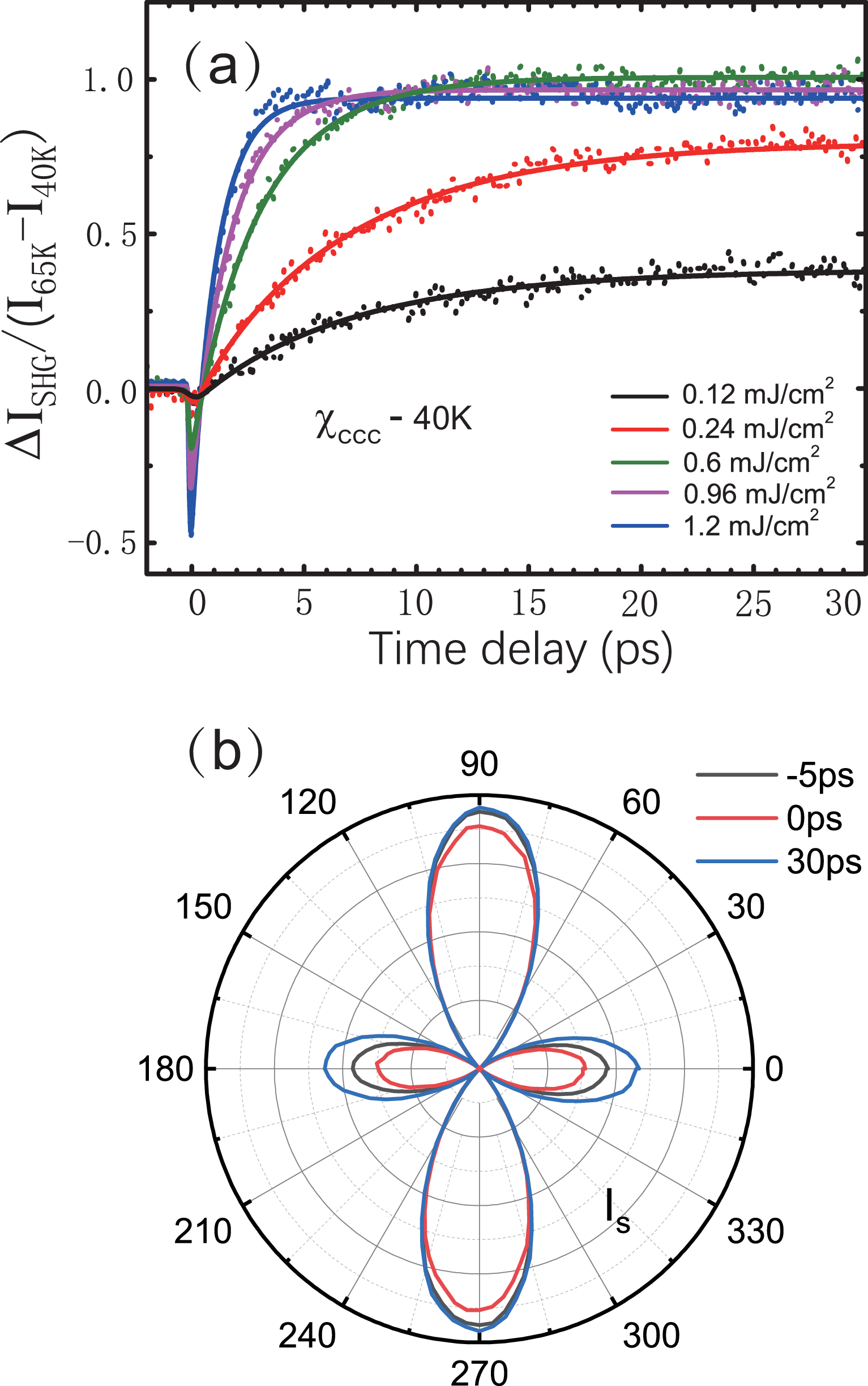}
	\caption{(a) Transient dynamics of the SHG change (normalized to $I_{65\mathrm{K}} - I_{40\mathrm{K}}$) at various pump fluences. The solid lines are fits using Eq. (1). (b) Polarization dependence of the SHG signal in the s-out configuration at -5 ps (before pump), 0 ps (at the dip), and 30 ps, measured at 40 K with a pump fluence of 1.2 mJ/cm$^2$.}\label{Fig:3}
\end{figure}

\begin{figure}[htbp]
	\centering
	\includegraphics[width=7cm]{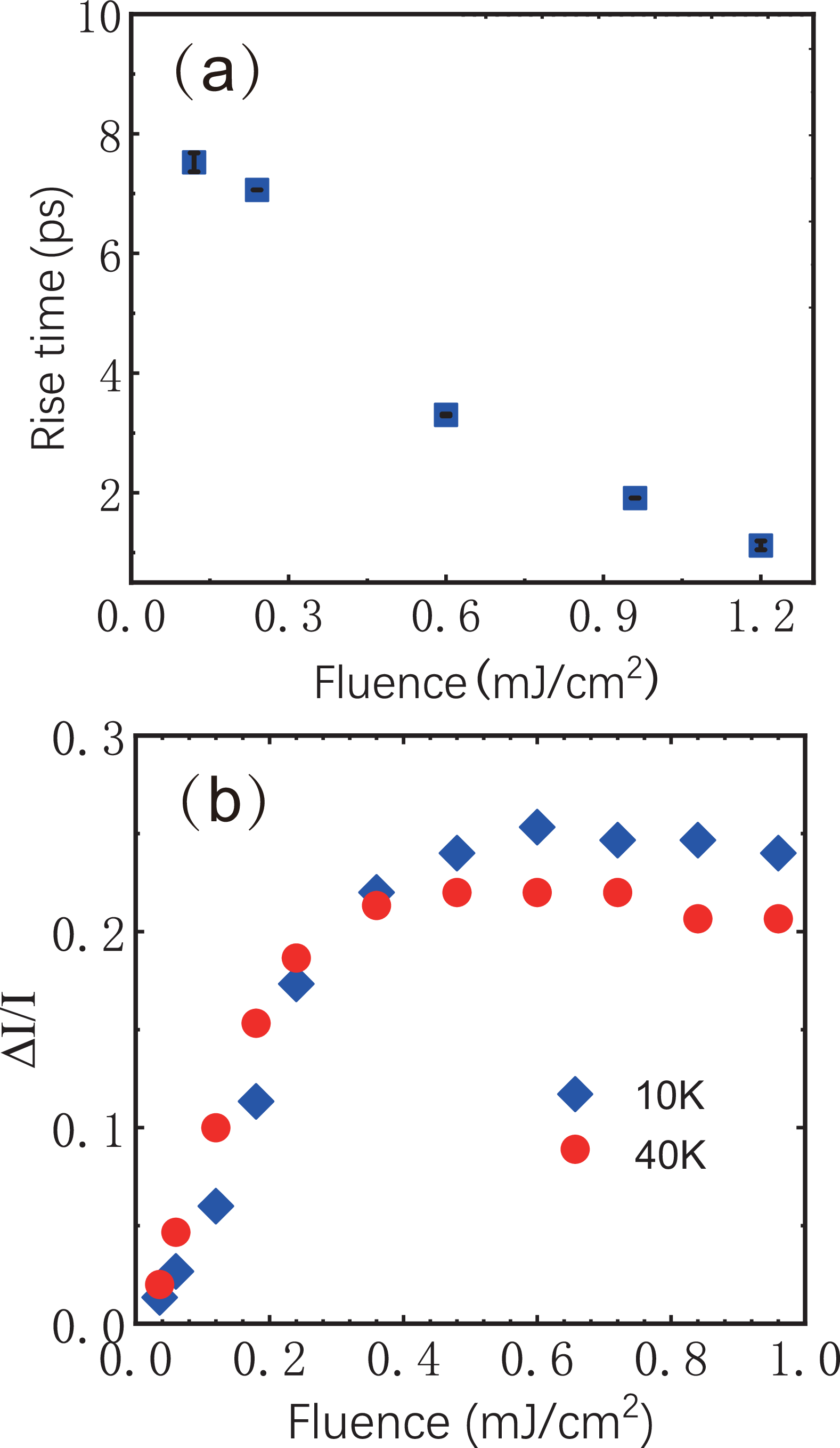}
	\caption{(a) Fluence dependence of the time constant $\tau_m$ obtained from the fits in Fig. 3(a). (b) Fluence dependence of the relative SHG change ($\Delta I/I$) at 30 ps (on the plateau) at 10 K and 40 K.}\label{Fig:4}
\end{figure}


Since only $\chi_{ccc}^{(2)}$ is sensitive to the magnetic order induced polarization, the time-resolved SHG measurement is mainly focused on this tensor element using the s-in/s-out configuration. Figure \ref{Fig:3}(a) shows the time evolution of the pump-induced SHG change at 40 K under various pump fluences. The transient response contains two distinct features: a sharp decrease followed by a fast recovery ($\tau_1 < 1$ ps), and a subsequent slow rise to a plateau with a time constant $\tau_m$ that decreases from $\sim 8$ ps at low fluence to $\sim 2$ ps at saturation. The fast component is observed for all tensor elements, while the slow component is specific to $\chi_{ccc}^{(2)}$. To extract the time scales quantitatively, we fit the data using

\begin{equation}
\frac{\Delta I}{I} = \Theta(t) \left[ a_1 \exp(-t/\tau_1) + a_m (1 - \exp(-t/\tau_m)) + c \right],
\end{equation}

where $\Theta(t)$ is the Heaviside step function with a 35 fs rise time, $a_1$ and $a_m$ are the amplitudes of the fast and slow components, respectively. The fitting curves are shown as solid lines in Fig. \ref{Fig:3}(a). The fluence dependence of the fitted time constant $\tau_m$ is presented in Fig. \ref{Fig:4}(a), while Fig. \ref{Fig:4}(b) shows the fluence dependence of the SHG signal change at 30 ps (on the plateau) at 10 K and 40 K.

We attribute the fast component ($\tau_1 < 1$ ps) to the ultrafast modulation of the spontaneous polarization by photo-excited carriers. The laser pulse excites electrons from the valence band to the conduction band, where delocalized carriers screen the dipole moments and temporarily reduce the SHG intensity. The polarization recovers as the carriers decay via electron-phonon scattering. This process is present in all tensor components, consistent with a transient modulation of the spontaneous polarization. This ultrafast polarization modulation process is an effective way to generate terahertz emission \cite{Shi_2020}.

The slow component, in contrast, is specific to $\chi_{ccc}^{(2)}$, immediately suggesting that it originates from the magnetically induced polarization identified by the static SHG measurements. To confirm this assignment, we performed polar-angle scans of the s-out SHG signal at -5 ps (before pump), 0 ps (at the dip), and 30 ps (on the plateau), as shown in Fig. \ref{Fig:3}(b). At -5 ps, the pattern matches the static SHG. At 0 ps, all tensor components are uniformly reduced, consistent with the fast polarization screening. At 30 ps, only the $\chi_{ccc}^{(2)}$-related lobe is significantly enhanced, recovering toward the high-temperature (paramagnetic) pattern. This indicates that the slow component corresponds to the photo-induced melting of the magnetic order and the associated erasure of the exchange-striction-induced polarization. At saturation, the SHG signal fully recovers to the value measured above $T_N$, confirming the nearly complete suppression of the magnetic order. The saturation fluence is only $\sim 0.4$ mJ/cm$^2$, about one order of magnitude lower than previously reported for demagnetization in insulating multiferroics \cite{PhysRevB.92.184429,PhysRevB.94.184429,PhysRevB.97.125149}.

We also examined the polarization dependence of the pump pulse. The SHG response is independent of the pump polarization: circularly polarized, linearly polarized with different orientations, and 2 $\mu$m pumping (using an optical parametric amplifier) all produced essentially the same result (see Supplemental Material). This contrasts with the circular-polarization-selective Kerr rotation reported previously in Fe$_2$Mo$_3$O$_8$ \cite{PhysRevX.9.031038}, where circularly polarized light was found to induce a net magnetization by selectively flipping spins in the antiferromagnetic sublattices. In our SHG measurements, which probe the lattice polarization rather than the magnetization directly, both circular and linear pumping efficiently suppress the exchange-striction-induced polarization. This suggests that melting the primary antiferromagnetic order can be achieved by photoexcitation regardless of the pump polarization, while the generation of net magnetization may require additional chiral selectivity.

Our time-resolved SHG measurements demonstrate that the magnetically induced polarization in Fe$_2$Mo$_3$O$_8$ can be efficiently suppressed by ultrafast optical pumping, with a fluence as low as $\sim 0.4$ mJ/cm$^2$. However, the recovery of the polar order occurs on a timescale beyond 100 ps, indicating that the photo-induced melting of the magnetic order is not followed by an equally fast restoration. This asymmetry between melting and recovery times suggests that while the excitation and initial quenching of the magnetic order are ultrafast, the subsequent relaxation involves slower processes such as lattice thermalization and magnetic domain recombination. To achieve faster recovery and enable repetitive switching, alternative excitation pathways may be required. In particular, intense terahertz pulses, which can resonantly drive low-energy modes such as phonons or magnons, may offer a more direct and efficient route to manipulate the exchange-striction-induced polarization without significant lattice heating.

In summary, by combining static and time-resolved SHG measurements, we have investigated the ultrafast dynamics of the magnetically induced polarization in Fe$_2$Mo$_3$O$8$. Static SHG reveals that only the $\chi^{(2)}_{ccc}$ tensor element exhibits a pronounced anomaly at $T_N$, identifying it as the nonlinear susceptibility component coupled to the magnetically induced polarization. Guided by this identification, time-resolved SHG shows that this magnetically induced polarization can be efficiently suppressed within several hundred femtoseconds upon optical excitation, reaching saturation at a remarkably low fluence of $\sim0.4$ mJ/cm$^2$, followed by a biexponential recovery. More broadly, this work demonstrates that combining static and time-resolved SHG provides a tensor-selective strategy for identifying and tracking magnetically induced polarization dynamics in multiferroic materials.

\section{Acknowledgments}
We would like to thank Profs. S. Dong, S. W. Cheong, J. Demsar for helpful discussions. This work was supported by National Natural Science Foundation of China (No. 12488201), the National Key Research and Development Program of China (No. 2024YFA1408701).

\bibliographystyle{apsrev4-1}
\bibliography{Fe2Mo3O8_SHGdatabase}

\newpage
\onecolumngrid

\renewcommand{\theequation}{S\arabic{equation}}
\setcounter{equation}{0}
\renewcommand{\thefigure}{S\arabic{figure}}
\setcounter{figure}{0}

\begin{center}
    {\Large \textbf{Supplemental Material: Identifying and tracking magnetically induced polarization in Fe$_2$Mo$_3$O$_8$ by static and time-resolved second harmonic generation}}
    
    \vspace{0.3cm}
   
\end{center}

\subsection*{Polarization-resolved SHG fitting}

We present details of the polar pattern fitting. The second-order polarization $P_i(2\omega)$ is related to the electric field $E_j(\omega)$ by

\begin{equation}
P_i = \epsilon_0 \sum_{j,k} d_{ijk} E_j E_k,
\end{equation}

where $d_{ijk}$ is the nonlinear susceptibility tensor. For SHG, $E_j E_k = E_k E_j$, so $d_{ijk} = d_{ikj}$. In the contracted notation, the subscripts 23 and 32 are substituted by 4, 31 and 13 by 5, and 12 and 21 by 6. Equation (S1) can be written as

\begin{equation}
\begin{pmatrix}
P_1 \\ P_2 \\ P_3
\end{pmatrix}
=
\begin{pmatrix}
d_{11} & d_{12} & d_{13} & d_{14} & d_{15} & d_{16} \\
d_{21} & d_{22} & d_{23} & d_{24} & d_{25} & d_{26} \\
d_{31} & d_{32} & d_{33} & d_{34} & d_{35} & d_{36}
\end{pmatrix}
\begin{pmatrix}
E_1^2 \\ E_2^2 \\ E_3^2 \\ 2E_2E_3 \\ 2E_1E_3 \\ 2E_1E_2
\end{pmatrix}.
\end{equation}

In our experimental geometry (Fig. \ref{Fig:1}(b)), the lab axes 1, 2, 3 correspond to the crystal axes $a$, $b$, $c$, respectively. The light is near-normal incident on the ac-plane, so the electric field components are

\begin{equation}
E_1 = E \sin\theta, \quad E_2 = 0, \quad E_3 = E \cos\theta,
\end{equation}

where $\theta$ is the polarization angle measured from the $c$-axis.

For point group 6mm (space group P6$_3$mc), the nonzero tensor components are $d_{33}$ and $d_{15} = d_{31} = d_{32} = d_{26}$. Equation (S2) then reduces to

\begin{equation}
\begin{pmatrix}
P_1 \\ P_2 \\ P_3
\end{pmatrix}
=
\begin{pmatrix}
0 & 0 & 0 & 0 & d_{15} & 0 \\
0 & 0 & 0 & 0 & 0 & d_{26} \\
d_{31} & d_{32} & d_{33} & 0 & 0 & 0
\end{pmatrix}
\begin{pmatrix}
E^2 \sin^2\theta \\ 0 \\ E^2 \cos^2\theta \\ 0 \\ 2E^2 \sin\theta\cos\theta \\ 0
\end{pmatrix}.
\end{equation}

Carrying out the multiplication gives

\begin{equation}
\begin{pmatrix}
P_1 \\ P_2 \\ P_3
\end{pmatrix}
=
\begin{pmatrix}
2 d_{15} E^2 \sin\theta \cos\theta \\
0 \\
d_{31} E^2 \sin^2\theta + d_{33} E^2 \cos^2\theta
\end{pmatrix}.
\end{equation}

Thus, the SHG intensities for s-out and p-out polarizations are

\begin{equation}
I_s \propto \left| \chi_{ccc}^{(2)} \cos^2\theta + \chi_{caa}^{(2)} \sin^2\theta \right|^2,
\end{equation}

\begin{equation}
I_p \propto \left| 2 \chi_{aca}^{(2)} \sin 2\theta \right|^2,
\end{equation}

which are the expressions used to fit the polar patterns in Fig. \ref{Fig:2}(a) and (b).

In the pump-probe measurements, the probe beam is reflected from the sample at a small incident angle rather than at normal incidence. The incident plane is horizontal (the $ab$-plane). For the $i$-type tensor, the $a$- and $b$-axes are equivalent, so the small projection of $E_a$ onto $E_b$ does not affect the measured SHG response. The results are therefore equivalent to the normal-incidence case.

\subsection*{Pump at 2 $\mu$m}

We also performed pump-probe measurements using a pump wavelength of 2 $\mu$m, generated by an optical parametric amplifier, to excite different electron transitions. Figure S1 shows the time traces of the SHG signal corresponding to $\chi_{ccc}^{(2)}$ under 2 $\mu$m pumping. The fast spike-like component disappears, while the slow component associated with the magnetic order melting remains qualitatively unchanged. The result is independent of the pump polarization. The fluence of the 2 $\mu$m pump pulse was above 1.5 mJ/cm$^2$, significantly higher than that of the 800 nm pump due to weaker absorption at 2 $\mu$m.

\begin{figure}[htbp]
	\centering
	\includegraphics[width=8cm]{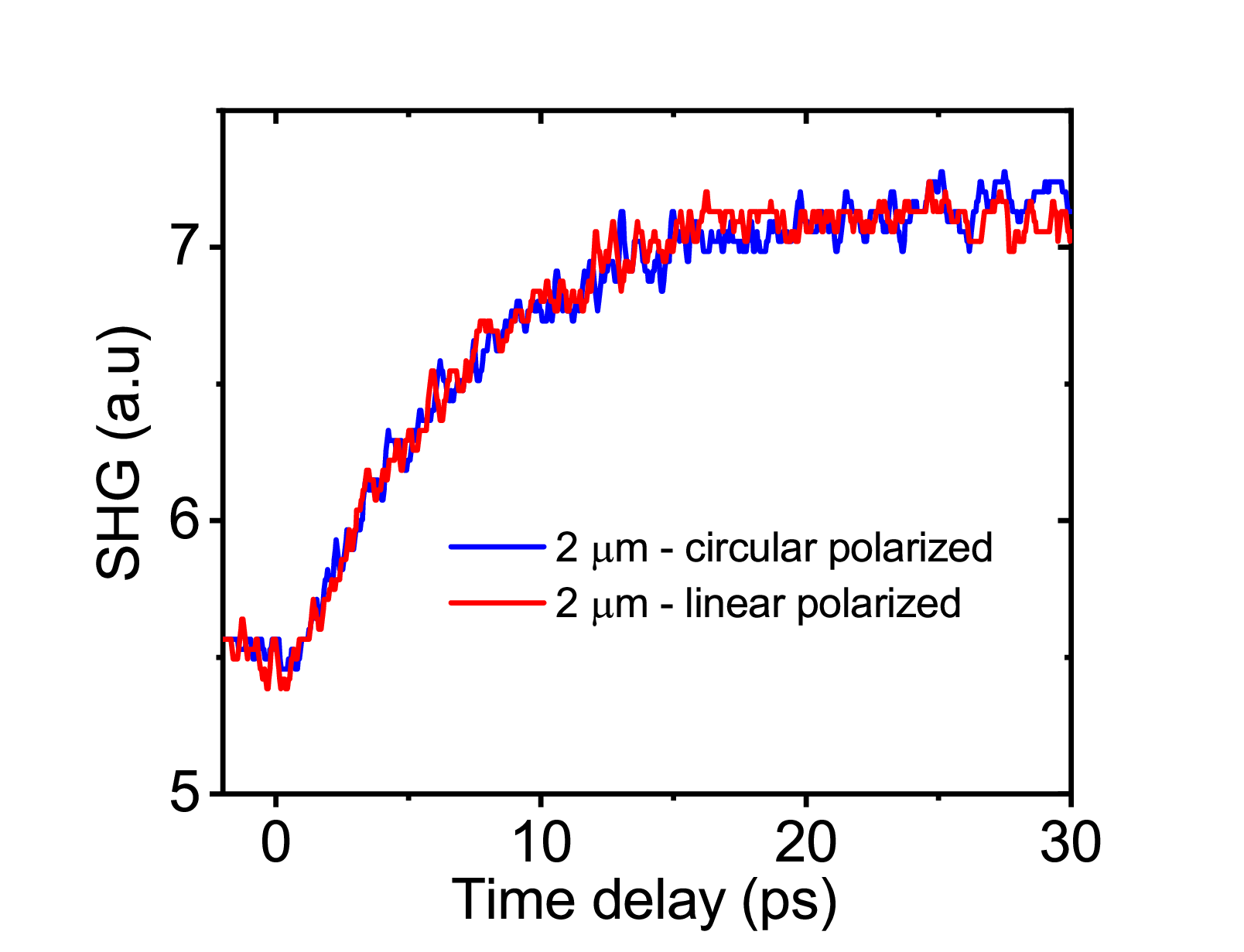}
	\caption{Time traces of the SHG signal corresponding to $\chi_{ccc}^{(2)}$ under 2 $\mu$m pumping. Linear and circular polarizations produce similar effects.}
	\label{Fig:S1}
\end{figure}

\end{document}